\documentclass[twocolumn,showpacs,aps,groupedaddress,prd,eqsecnum]{revtex4}
\usepackage{amsfonts,latexsym,eucal}
\usepackage[dvips]{graphicx}
\usepackage{epsfig}
\usepackage{color}

\def\beq{\begin{equation}}
\def\eeq{\end{equation}}
\def\bea{\begin{eqnarray}}
\def\eea{\end{eqnarray}}
\def\nn{\nonumber\\}
\def\pa{\partial}
\def\ra{\rightarrow}
\def\bp{\mbox{\boldmath$\phi$}}

\def\bxi{\mbox{\boldmath$\xi$}}
\def\bpi{\mbox{\boldmath$\varpi$}}

\def\tt{\tilde{t}}
\def\tr{\tilde{r}}
\def\tp{\tilde{\phi}}
\def\to{\tilde{\omega}}
\def\tE{\tilde{E}}
\def\tQ{\tilde{Q}}

\begin{document}
\draft

\thispagestyle{empty}

\title{What happens to Q-balls if $Q$ is so large?}
\author{Nobuyuki Sakai}
\email{nsakai@e.yamagata-u.ac.jp}
\affiliation{Department of Education, Yamagata University, Yamagata 990-8560, Japan}
\author{Takashi Tamaki}
\email{tamaki@ge.ce.nihon-u.ac.jp}
\affiliation{Department of Physics, General Education, College of Engineering, 
Nihon University, Tokusada, Tamura, Koriyama, Fukushima 963-8642, Japan}

\begin{abstract}
In the system of a gravitating Q-ball, there is a maximum charge $Q_{{\rm max}}$ inevitably, 
while in flat spacetime there is no upper bound on $Q$ in typical models such as the Affleck-Dine model.
Theoretically the charge $Q$ is a free parameter, and phenomenologically it could increase by charge accumulation.
We address a question of what happens to Q-balls if $Q$ is close to $Q_{{\rm max}}$.
First, without specifying a model, we show analytically that inflation cannot take place in the core of a Q-ball, contrary to the claim of previous work.
Next, for the Affleck-Dine model, we analyze perturbation of equilibrium solutions with $Q\approx Q_{{\rm max}}$ by numerical analysis of dynamical field equations.
We find that the extremal solution with $Q=Q_{{\rm max}}$ and unstable solutions around it are ``critical solutions", which means the threshold of black-hole formation.
\end{abstract}

\pacs{04.25.dc, 04.40.-b, 11.27.+d, 95.35.+d}
\maketitle
 
\section{Introduction}

In a pioneering work by Friedberg {\it et al}.\ in 1976 \cite{FLS76}, nontopological solitons were introduced in a model with a U(1)-symmetric complex scalar field coupled to a real scalar field.
In contrast with topological defects, they are stabilized by a global U(1) charge, and their energy density is localized in a finite space region without gauge fields.
In 1985 Coleman showed such solitons exist in a simpler model with an SO(2) [viz.\ U(1)] symmetric scalar field only, and called them Q-balls \cite{Col85}.

Q-balls have attracted much attention in particle cosmology since Kusenko pointed out that they can exist in all supersymmetric extensions of the standard model \cite{Kus97b-98}.
Specifically, Q-balls can be produced efficiently in the Affleck-Dine (AD) mechanism \cite{AD} and could be responsible for baryon asymmetry \cite{SUSY} and dark matter \cite{SUSY-DM}.
Q-balls can also influence the fate of neutron stars \cite{Kus98}.
Since Q-balls are supposed to be microscopic objects,
equilibrium solutions and their stability have been intensively studied in flat spacetime \cite{stability}.
It was shown that catastrophe theory is a useful tool for stability analysis of Q-balls \cite{SS}.

If Q-balls are so large or so massive, on the other hand, their size becomes astronomical and their gravitational effects are remarkable \cite{Grav-Q,multamaki}.
Such gravitating Q-balls, or Q-stars, are analogous to boson stars \cite{boson-review}.
While Q-balls exist even in flat spacetime, boson stars are supported by gravity and nonexistent in flat spacetime.
Multamaki and Vilja showed that the size of Q-balls is bounded above due to gravity \cite{multamaki}.
Becerril {\it et al}.\ studied evolution of unstable solutions by numerical analysis of dynamical field equations \cite{BBGN}.

In our previous work \cite{TS1,TS3,TS24} we studied equilibrium solutions and their stability for several models using catastrophe theory to understand a comprehensive picture of flat Q-balls, gravitating Q-balls and boson stars.
One of the main results is summarized in TABLE I.
\begin{itemize}
\item In flat spacetime, as long as the absolute minimum of the potential $V(\phi)$ is located at $\phi=0$, there is no upper bound on charge.
If we take self-gravity into account, however, there is maximum charge $Q_{{\rm max}}$ 
(or maximum  energy $E_{{\rm max}}$),  at which stability changes, regardless of models.
\item In flat spacetime, in some models such as the AD gauge-mediated model \cite{AD}, there is nonzero minimum charge $ Q_{{\rm min}}$, where Q-balls with $Q<Q_{{\rm min}}$ are nonexistent.
If we take self-gravity into account, however, there exist stable Q-balls with arbitrarily small mass and charge.
\end{itemize}
The above properties of gravitating solitons hold for general models of Q-balls and boson stars 
as long as the leading order term of the potential is a positive mass term in its Maclaurin series.

\begin{table}
\begin{tabular}{|c|c|c|}\hline
& $Q_{{\rm max}}$ & $Q_{{\rm min}}$ \\\hline
flat Q-balls & $\infty$ or finite & finite or 0 \\\hline
gravitating Q-balls & finite & 0 \\\hline
boson stars & finite & 0 \\\hline
\end{tabular}
\caption{Comparison of maximum and minimum charge (or gravitational mass) for flat Q-balls, gravitating Q-balls and boson stars.}
\end{table}

Theoretically the charge $Q$ is a free parameter, and phenomenologically it could 
increase by charge accumulation.
Therefore, a natural question could arise: what happens to Q-balls if $Q$ is close to $Q_{{\rm max}}$?
In this paper we shall investigate what happens if we give perturbations to equilibrium 
Q-balls with $Q\approx Q_{{\rm max}}$.

In connection with this question, Matsuda \cite{Mat} claimed that inflation occurs in the core of a Q-ball if $Q$ is large enough.
He assumed a kind of hybrid potential. In his scenario inflation takes place when the gravity-mediated term,
\beq\label{AD}
V(\phi)\equiv \frac{m^2}{2}\phi^2\left[1+K\ln \left(\frac{\phi}{M}\right)^2\right],
\eeq
dominates, where $m$ is the gravitino mass, $K$ term a one-loop correction, and $M$ the renormalization scale.
If this scenario is true, it would have an important implication for inflationary models.
Therefore, the present study is also important as a close examination of this scenario.

This paper is organized as follows.
In Sec. II we review previous results of gravitating Q-balls in the AD gravity-mediated model.
In Sec. III we make analytic discussions on general properties of Q-ball equilibrium solutions.
In Sec. IV we discuss whether Q-ball inflation can take place.
In Sec. V we present dynamical field equations and our computing method.
In Sec. VI, by analyzing the dynamical field equations, we investigate what happens if $Q$ is so large in the AD gravity-mediated model.
Sec. VII is devoted to concluding remarks. 

\section{Gravitating Q-balls in the Affleck-Dine potential}

To begin with, we review previous results of equilibrium solutions of gravitating Q-balls in the AD gravity-mediated model  (\ref{AD})  \cite{TS3}.
Consider an SO(2)-scalar field $\bp=(\phi_1,\phi_2)$ coupled to Einstein gravity.
The action is given by
\beq\label{Sg}
{\cal S}=\int d^4x\sqrt{-g}
\left\{{{\cal R}\over 16\pi G}-\frac12g^{\mu\nu}\pa_{\mu}\bp\cdot\pa_{\nu}\bp-V(\phi) \right\},
\eeq
where $\phi\equiv\sqrt{\phi_1^2+\phi_2^2}$.
We assume a spherically symmetric and static spacetime, 
\beq\label{metric}
ds^2=-\alpha^2(r)dt^2+A^2(r)dr^2+r^2(d\theta^2+\sin^2\theta d\varphi^2),
\eeq
and homogeneous phase rotation,
\beq\label{phase}
\bp (t,r)=\phi(r)(\cos\omega t,\sin\omega t).
\eeq

\begin{figure}[htbp]
\psfig{file=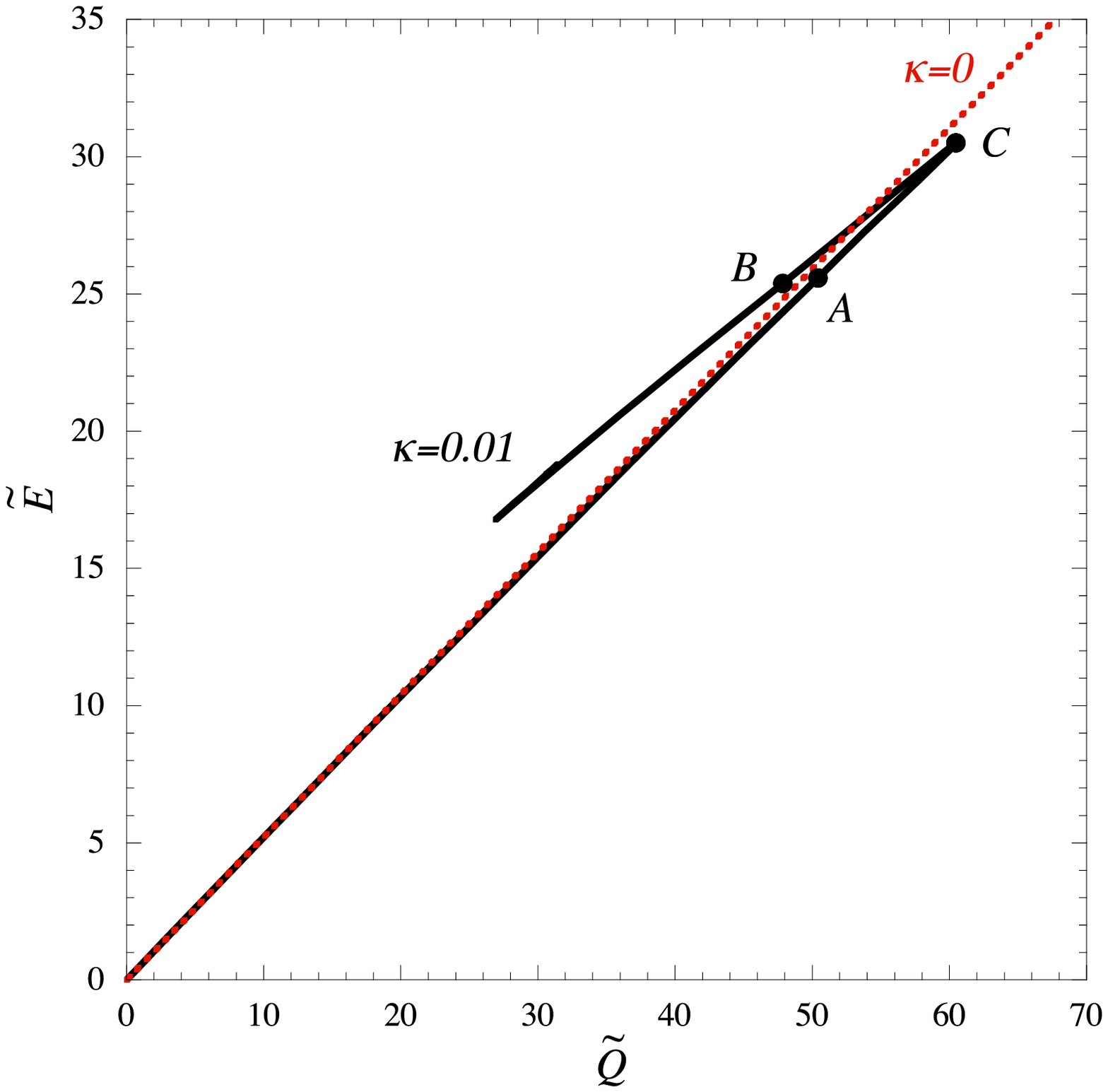,width=3in}
\caption{
Existing domain of equilibrium solutions for $K=-0.01$ in the $Q$-$E$ space.
We choose $\kappa=0,~0.01$.}
\end{figure}

Then the field equations become
\bea\label{Gtt}
-{r A^3\over2}G^t_t&\equiv&A'+{A\over2r}(A^2-1) \nonumber \\
&=&{4\pi G}r A^3\left({{\phi'}^2\over2A^2}
+{\omega^2\phi^2\over2\alpha^2}+V\right),
\\\label{Grr}
{r\alpha\over2}G_{rr}&\equiv&\alpha'+{\alpha\over2r}(1-A^2) \nonumber \\
&=&{4\pi G}r\alpha A^2
\left({{\phi'}^2\over2A^2}+{\omega^2\phi^2\over2\alpha^2}-V\right),
\\\label{Box}
{A^2\phi\over\phi}\Box\phi&\equiv&
\phi''+\left(\frac2r+{\alpha'\over\alpha}-{A'\over A}\right)\phi'
+\left({\omega A\over\alpha}\right)^2\phi \nonumber \\
&=&A^2{dV\over d\phi}.
\eea
The boundary conditions are given by
\bea\label{BC}
&&\phi'(0)=\phi(\infty)=A'(0)=\alpha'(0)=0,\nn
&&A(0)=\alpha(\infty)=1.
\eea

The charge and the Hamiltonian energy are defined by \cite{TS1}
\bea\label{Q}
Q&\equiv&\int d^3x\sqrt{-g}g^{0\nu}(\phi_1\pa_\nu\phi_2-\phi_2\pa_\nu\phi_1),
\nn
E&\equiv&\lim_{r\ra\infty}{r^2\alpha'\over2GA}={M_S\over2},
\label{H}\eea
where $M_S$ is the Schwarzschild mass.
To perform numerical calculations, we rescale the relevant quantities as
\bea
&&\tp\equiv\frac{\phi}{M},~~\tt\equiv mt,~~ \tr\equiv mr~~\to\equiv\frac{\omega}{m},~~
\kappa\equiv GM^2, \nn
&&
\tQ\equiv{m^2\over M^2}Q,~~\tE\equiv{m\over M^2}E.
\eea

Figure 1 shows existing domain of equilibrium solutions for $K=-0.01$ in the $Q$-$E$ space.
In the case of flat spacetime ($\kappa=0$) no upper bound on $Q$ or $E$ appears and 
all the solutions in this figure are stable~\cite{foot}.
In the case of $\kappa=0.01$, on the other hand, there appears a cusp $A$, which corresponds to 
$Q_{\rm max}$ and $E_{\rm max}$. 
The lower branch represents stable solutions, while the upper branch unstable ones.

\begin{figure}[htbp]
\psfig{file=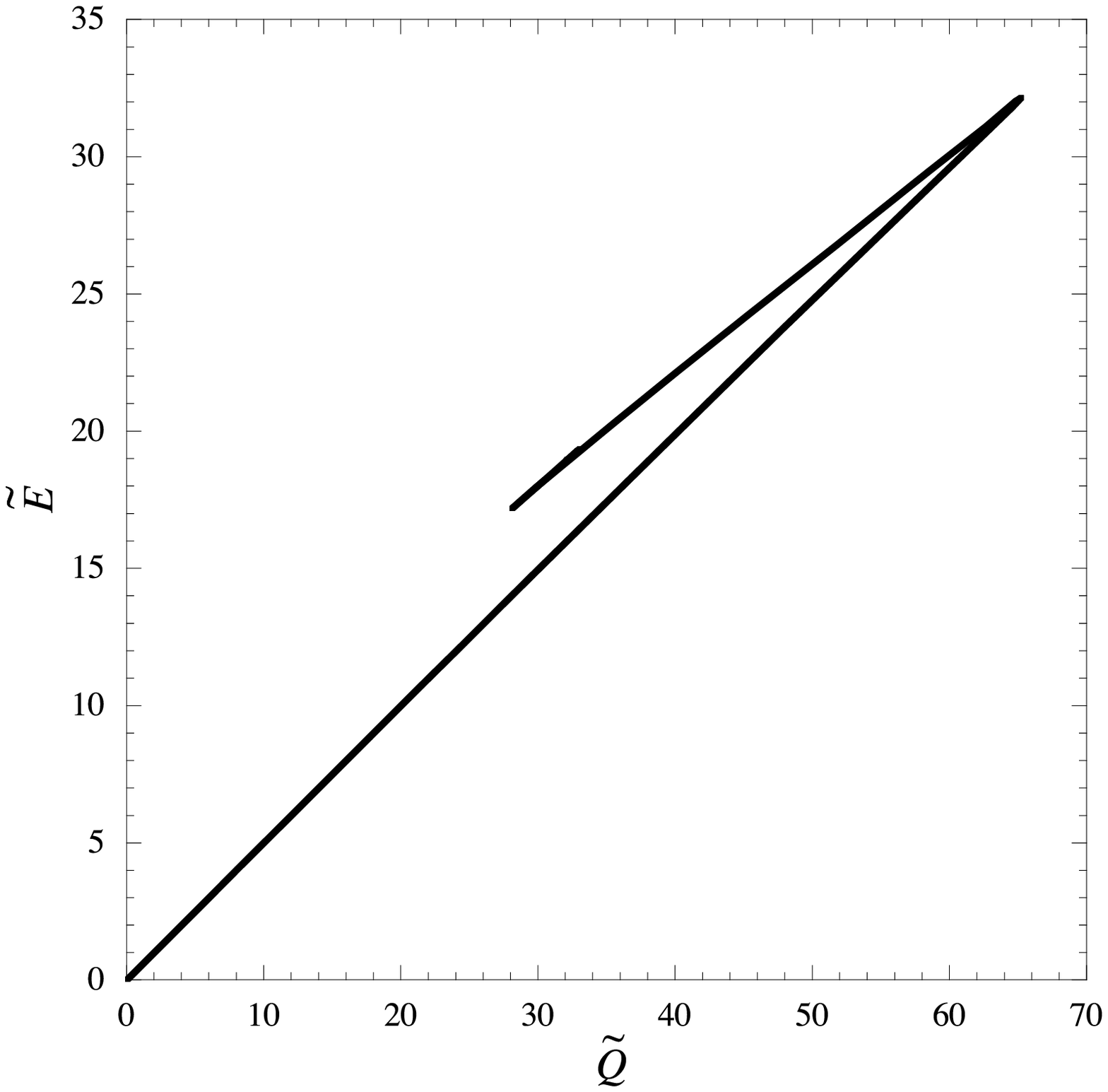,width=3in}
\caption{
Existing domain of equilibrium solutions for $K=0$ in the $Q$-$E$ space. 
We choose $\kappa=0.01$.}
\end{figure}

In flat spacetime equilibrium solutions exist only if $K<0$.
If we take self-gravity into account, however, equilibrium solutions also exist even if $K\ge0$  \cite{TS3}.
The solutions for $K=0$ have been known as mini-boson stars~\cite{boson-review}. 
Figure 2 shows existing domain of these equilibrium solutions in the $Q$-$E$ space.
The result is similar to that for $K<0$.
There appears a cusp, which corresponds to $Q_{\rm max}$ and $E_{\rm max}$, due to gravity.
The lower branch represents stable solutions, while the upper branch unstable ones. 

One may wonder what happens to equilibrium solutions if $\kappa$ is so large.
In the case of topological defects, static solutions are nonexistent if the vacuum expectation value of the Higgs field is close to the Planck mass, and the defects expand exponentially instead \cite{GRLV}.
By analogy with this topological inflation, one may conjecture that Q-ball equilibrium solutions are nonexistent if $\kappa$ is larger than some critical value of order one.

To examine this conjecture, for the case of $K=-0.01$, we analyze equilibrium solutions with 
$\kappa=0.1,~1,~10,~100,~1000$ and 10000, too.
We show the existing domains of the equilibrium solutions in Fig.\ 3(a).
Contrary to the above expectation, 
we find that equilibrium solutions exist even if $\kappa\gg1$.
Furthermore, Fig.\ 3(a) indicates
\beq\label{kappaE}
\kappa \tE_{\rm max}={\rm constant}, 
\eeq
where $\tE_{\rm max}$ denotes the maximum energy  for each $\kappa$.
Because the Schwarzschild mass of a Q-ball is given by $M_S=2E$ \cite{TS1}, 
the relation (\ref{kappaE}) means $GM_S=$constant.
Therefore, as long as (\ref{kappaE}) is satisfied, we expect that static regular solutions can exist.

\begin{figure}[htbp]
\psfig{file=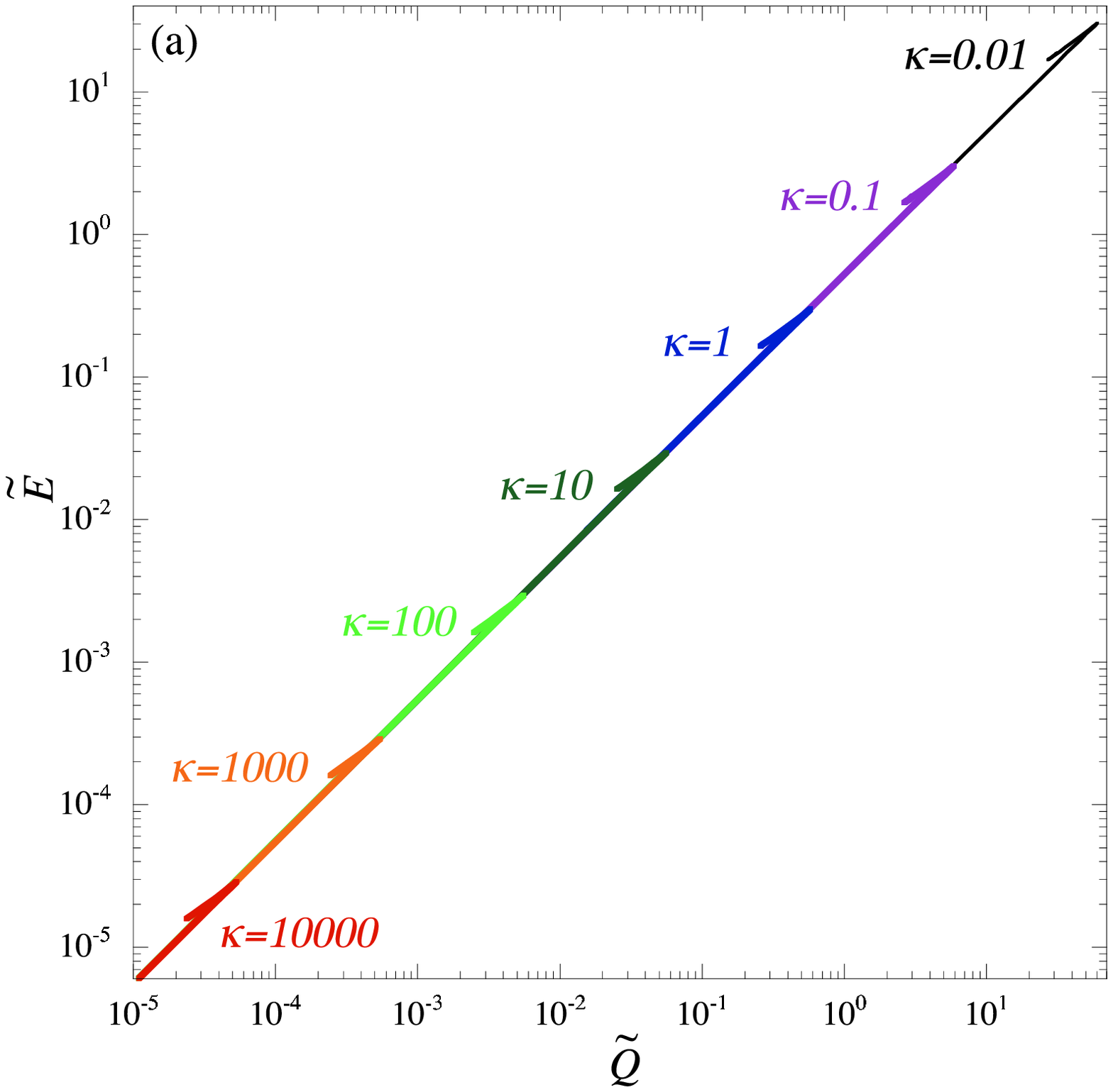,width=3in}
\psfig{file=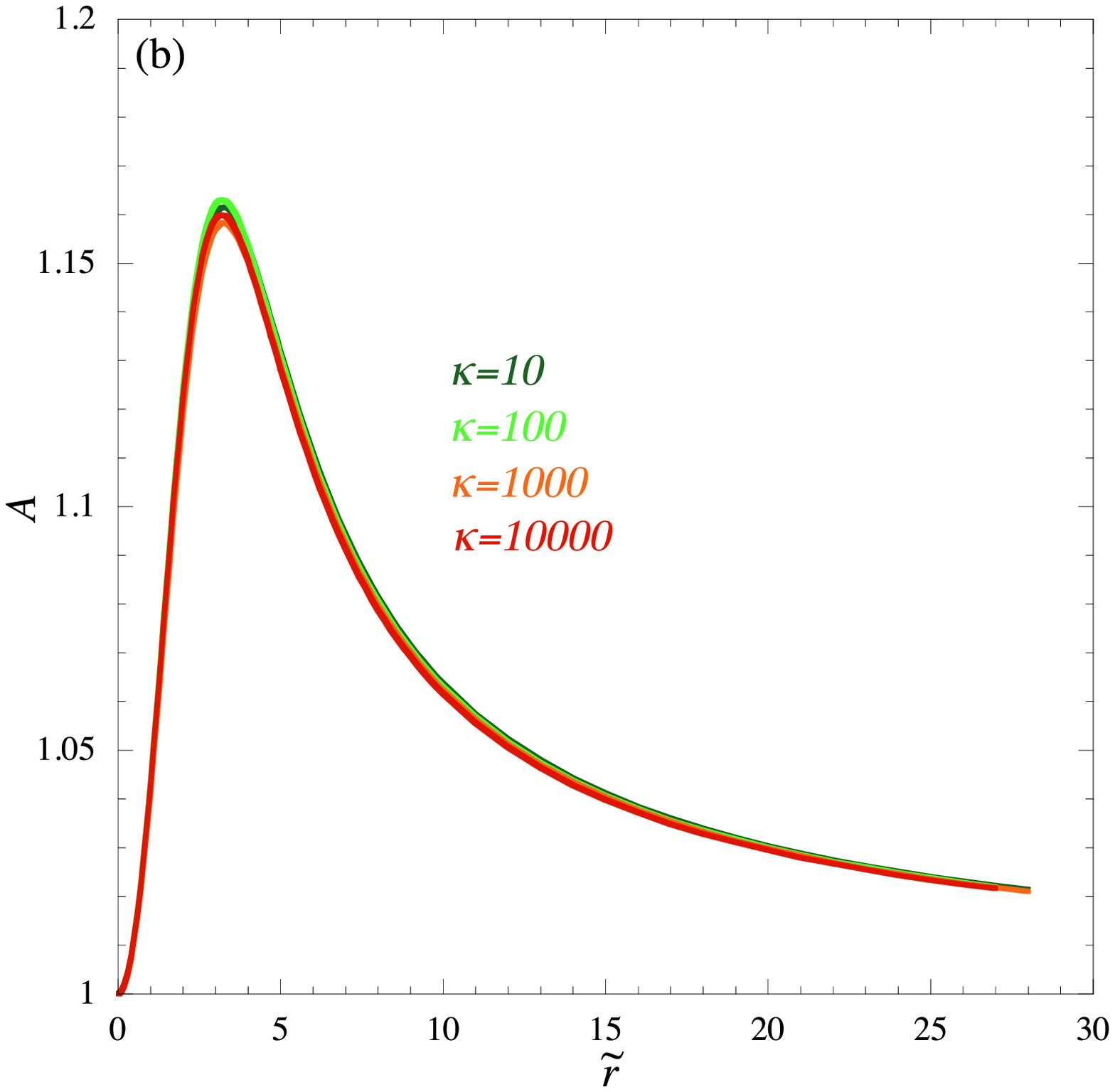,width=3in}
\caption{
Dependence of equilibrium solutions on $\kappa$.
We choose $K=-0.01$ and $\kappa=0.01,~0.1,~1,~10,~100,~1000$ and 10000.
(a) Existing domain of equilibrium solutions.
(b) The metric function $A(\tr)$ of the extremal solutions for each $\kappa$. }
\end{figure}

To confirm this argument, we present one of the metric functions $A(\tr)$ of the extremal solutions with $\tE_{\rm max}$ in Fig.\ 3(b).
All configurations of $A(\tr)$ are virtually the same, which is consistent with (\ref{kappaE}).
We therefore conclude that equilibrium solutions exist no matter how large $\kappa$ is.

\section{General properties of equilibrium solutions}

Our interest is perturbation of equilibrium solutions with $Q\approx Q_{{\rm max}}$.
Before analyzing their evolution, however, it is important to understand general properties of equilibrium solutions.

\begin{figure}[htbp]
\psfig{file=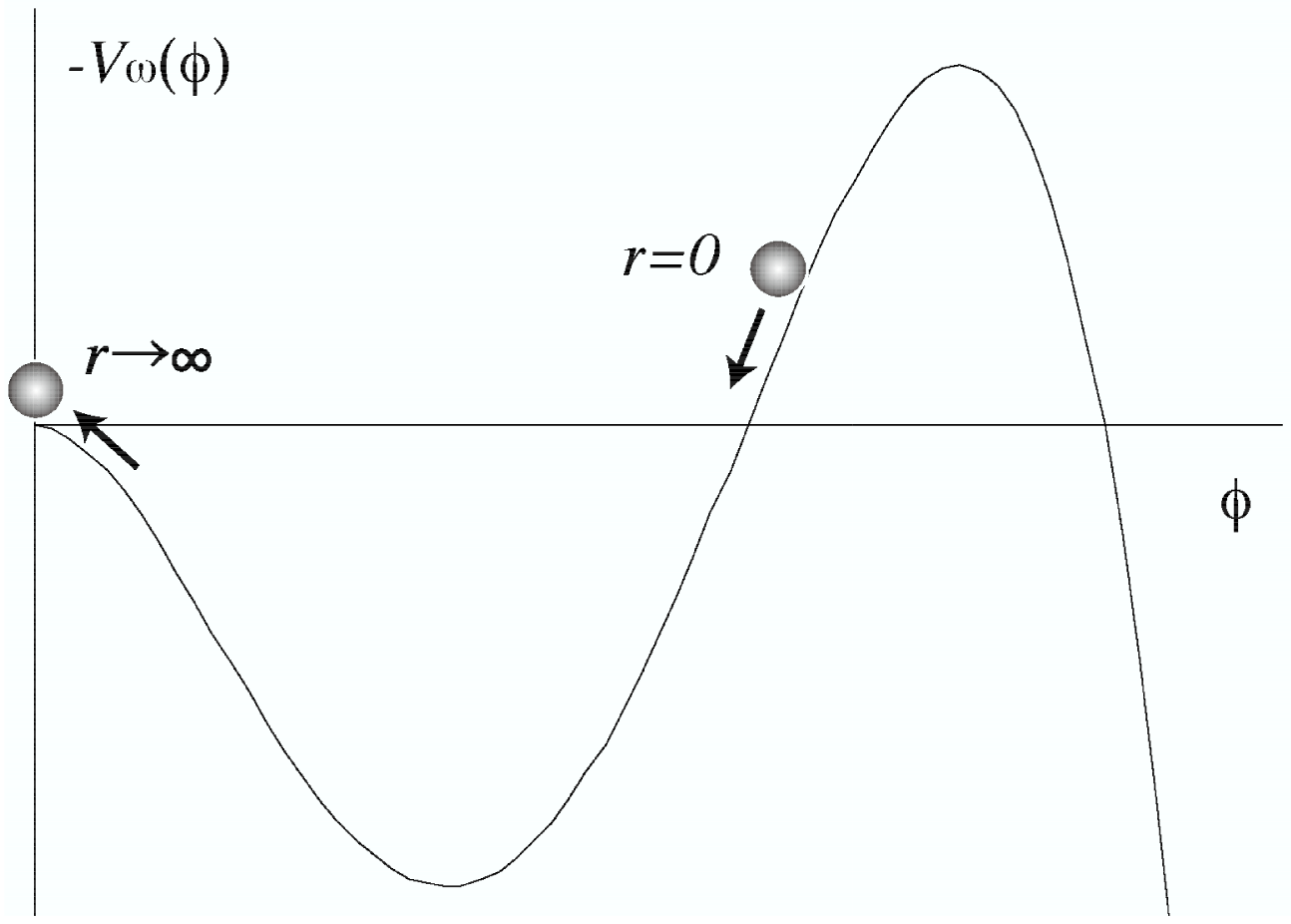,width=3in}
\caption{\label{Newton}
Interpretation of Q-ball solutions by analogy with a particle motion in Newtonian mechanics.}
\end{figure}

In the case of flat spacetime, the field equation is given by
\beq\label{FEqball}
\phi''+\frac2r\phi'+\omega^2\phi={dV\over d\phi}.
\eeq
This is equivalent to the field equation for a single static scalar field with a potential 
$V_{\omega}\equiv V-\omega^2\phi^2/2$.
If one regards the radius $r$ as \lq time\rq\ and the scalar amplitude $\phi(r)$ as \lq the position of
a particle\rq, one can understand Q-ball solutions in words of Newtonian mechanics, as shown in Fig.\ \ref{Newton}.
Equation (\ref{FEqball}) describes a one-dimensional motion of a particle under the conserved force due to the \lq potential\rq\ $-V_{\omega}(\phi)$ and the \lq time\rq-dependent friction $-2\phi'/r$.
If one chooses the \lq initial position\rq\ $\phi(0)$ appropriately, the static particle begins to roll 
down the potential slope, climbs up and approaches the origin over infinite time. 
Because of the energy loss by the friction term in Newtonian mechanics, one finds
\beq
-V_{\omega}={\omega^2\phi^2\over2}-V>0
~~~{\rm at}~~~ r\approx0.
\eeq
Dominance of the kinetic energy over the potential energy in the core is one of the important properties of Q-balls, as we shall discuss below.

We can extend the above argument to gravitating Q-balls by redefining the effective potential as 
$V_{\omega}\equiv{V}-{{\omega}^2}{\phi}^2/2\alpha^2$~\cite{TS3}.
In this case, because \lq the potential of a particle\rq\ is \lq time\rq-dependent, 
the above explanation in Newtonian mechanics cannot apply.
However, because $\alpha (r)$ is an increasing function, the \lq potential\rq\ $-V_{\omega}(\phi)$ decreases as \lq time\rq.
This means that some of the \lq mechanical energy\rq\ is lost by this variation of the \lq potential\rq\ as well as by the friction terms.
Therefore, we obtain the relation,
\beq\label{kindom}
-V_{\omega}={\omega^2\phi^2\over2\alpha^2}-V>0
~~~{\rm at}~~~ r\approx0.
\eeq
We have confirmed that the inequality (\ref{kindom}) holds for all our numerical solutions.

The condition (\ref{kindom}) is important because from it we can draw a general conclusion that the Q-ball core is attractive as follows.
Because the regularity condition at $r=0$ is given by (\ref{BC}), we can expand the metric functions as
\beq
\alpha(r)=\alpha(0)\left(1+\Phi(r)\right),~~~ A(r)=1+f(r),
\eeq
with the boundary conditions
\beq
\Phi(0)=\Phi'(0)=f(0)=f'(0)=0.
\eeq
In the vicinity of $r=0$ we can expand the geodesic equations and the Einstein equations 
up to first order of $\Phi$ and $f$, which yields
\bea\label{geodesic}
{1\over\alpha(0)^2}{d^2r\over dt^2}=-\Phi'&{\rm at}&r\approx0,
\\\label{Poisson}
\Phi''={8\pi G\over3\alpha(0)}\left({\omega^2\phi^2\over\alpha(0)^2}-V\right)
&{\rm at}&r\approx0.
\eea
Note that we have not introduced the weak-gravity approximation.

Equations (\ref{geodesic}) and (\ref{Poisson}) correspond to an equation of motion and Poisson equation, 
respectively, in Newtonian mechanics.
They lead to $\Phi''>0$ at $r\approx0$.
Furthermore, because of the boundary condition $\Phi'(0)=0$ , one finds $\Phi'>0$ at $r\approx0$, that is, the core region is attractive.

\section{Is Q-ball inflation possible?}

It was argued that inflation can take place in the core of a Q-ball if $Q$ evolves with time by absorbing other Q-balls and becomes large enough \cite{Mat}.
In the  last section, however,  we showed that Q-balls have attractive nature, which indicates that Q-ball inflation is improbable.
Here we shall discuss this issue more explicitly.

Inflation, or accelerated expansion of a local region, is defined by the following conditions.
\begin{enumerate}
\item Some local region is well-approximated by the Friedmann-Lemaitre-Robertson-Walker metric,
\beq\label{FLRW}
ds^2=-d\tau^2+a(\tau)^2\{d\chi^2+\chi^2(d\theta^2+\sin^2\theta\varphi^2)\}.
\eeq
\item $d^2a/d\tau^2>0$ in this region.
\item The volume of this local region increases.
\end{enumerate}
If any of these conditions are violated, we cannot say that inflation takes place.

The two expressions (\ref{metric}) and (\ref{FLRW}) look very different. In the case of topological inflation, however, the core of a topological defect is described by de Sitter spacetime,
\beq
\alpha^2=A^{-2}=1-H^2r^2,~~~
a\propto e^{H\tau}.
\eeq
which assures us of consistence between the two expressions.
If Q-balls also trigger inflation, the core of the equilibrium solutions should be well-approximated by (\ref{FLRW}).

Here, under the assumption that the first condition is satisfied for the equilibrium solutions, we discuss whether the second condition is satisfied.
The Einstein equations for the doublet scalar field yield
\beq\label{ddotaQ}
{d^2a\over d\tau^2}={8\pi Ga\over 3}\left(-\left|{d\bp\over d\tau}\right|^2+V\right).
\eeq
Although the relation between the two coordinate sets, $(t,r)$ and $(\tau,\chi)$, is not given explicitly, we find $\alpha dt=d\tau$ at the center ($r=\chi=0$). Then we can rewrite Eq.(\ref{ddotaQ}) as
\beq\label{ddotaQ2}
{d^2a\over d\tau^2}={8\pi Ga\over 3}\left(-{\omega^2\phi^2\over\alpha(0)^2}+V\right)
~~~{\rm at}~~~ r\approx0.
\eeq
Because $\omega^2\phi^2/\alpha(0)^2>2V$ from (\ref{kindom}), we find $d^2a/d\tau^2<0$ in the core.
Unless we give so large perturbation that $|d\bp/d\tau|^2/V$ is more than double, $d^2a/d\tau^2$ remains negative.

If we chose initial values of $\bp(x)$ and $d\bp/d\tau(x)$ arbitrary, the right-hand side of (\ref{ddotaQ}) might be positive. However, if the initial configuration is far from that of the equilibrium solution, such inflation cannot be called Q-ball inflation.

We should also note that we have not specified $Q$ and a potential type $V(\phi)$.
We therefore conclude that Q-ball inflation cannot take place by charge accumulation regardless of $Q$ and a potential type.

\section{Dynamical field equations and computing method}

Although we have shown that inflation cannot take place even if $Q>Q_{{\rm max}}$,
it is still unclear what happens in this case.
To address this question, we have to solve dynamical field equations.
Here, in preparation for this, we present dynamical field equations and our computing method.

We consider a spherically symmetric and dynamical spacetime, 
\beq\label{metric2}
ds^2=-\alpha^2(t,r)dt^2+A^2(t,r)dr^2+r^2(d\theta^2+\sin^2\theta d\varphi^2).
\eeq
Introducing dimensionless auxiliary variables, 
\bea\label{hojo}
&&\bpi\equiv{A\over\alpha}\dot{{\bp}}=(\varpi_1,~\varpi_2),~~~ 
\bxi\equiv{\pa\over\pa(r^2)}{\bp}=(\xi_1,~\xi_2),~~~\nn
&&a\equiv{A-1\over r^2},~~~{\rm with} ~~~ \dot{~}\equiv{\pa\over\pa t},
\eea
we write down the field equations derived from the action (\ref{Sg}) as
\bea\label{HC}
-{A^2}G^t_t&\equiv&{4\over A}{\pa A\over\pa(r^2)}+a(1+A) \nn
&=&8\pi G\left({\bpi\cdot\bpi\over2}+2r^2\bxi\cdot\bxi+A^2V\right),
\\ \label{MC}
{rA}G_{tr}&\equiv&\dot a=8\pi G\alpha\bpi\cdot\bxi,
\\ \label{Grr2}
r\alpha G_{rr}&\equiv&\alpha'-{r\alpha a(1+A)\over2} \nn
&=&{\kappa\over2}r\alpha\left({\bpi\cdot\bpi\over2}+2r^2\bxi\cdot\bxi-A^2V\right),
\\ \label{sfe}
{\alpha A}\Box\bp&\equiv&
-\dot{\bpi}+4r^2{\pa\over\pa(r^2)}\left({\alpha\bxi\over A}\right)
+{6\alpha\bxi\over A}\nn
&=&{\alpha A\bp\over\phi}{dV\over d\phi}.
\eea
We have regularized the dynamical equations at the center, in 
the sense that all of them contain no diverging term like $1/r $. 

As for initial conditions, we assume 
\beq
\dot a(0,r)=(\dot A=)0,
\eeq
and the perturbed field configuration, 
\beq\label{initial}
{\bp}(0,r)=(\phi_0(r)+\delta\phi(r),~0),~~\dot{\bp}(0,r)=(0,~\omega\phi_0(r)),
\eeq
where $\phi_0$ is an equilibrium solution and $\delta\phi$ is a small perturbation.
For definiteness we adopt
\beq\label{perturbation}
\delta\phi(r)=\delta\phi(0)\exp\left(-{r^2\over L^2}\right),
\eeq
where $L$ is a length parameter.
We rescale it as $\tilde{L}\equiv mL$ and set $\tilde{L}=2$ in Sec.\ VI.

To obtain initial values of $a$ (or $A$) and $\alpha$, 
we integrate (\ref{HC}) and (\ref{Grr2}) with respect to $r$. 
More precisely, to keep better precision, only at the initial time we introduce another auxiliary 
function as $Y\equiv A-1=ar^2$ and rewrite (\ref{HC}) as
\beq
Y'+{YA(A+1)\over2r}
={\kappa\over2}r A^3\left({\dot{{\bp}}\cdot\dot{{\bp}}\over2\alpha^2}
+{{\bp}'\cdot{\bp}'\over2A^2}+V\right).
\eeq

Our dynamical field variables are $\alpha,~a,~\bp,~\bxi$ and $\bpi$.
Among them the lapse function $\alpha$ is determined by integration of
(\ref{Grr2}) with respect to $r$ at each time step.
The rest of the dynamical variables are determined by integration of
(\ref{MC}) and (\ref{sfe}) together with
\beq
\dot{\bp}={\alpha\bpi\over A},~~~ 
\dot{\bxi}={\pa\over\pa (r^2)}\left({\alpha\bpi\over A}\right),
\eeq
which is given by the definition (\ref{hojo}).
The Hamiltonian constraint (\ref{HC}) is not solved except for the
initial values, but it is used to check numerical accuracy of the above 
time-integration.

To perform integration of the dynamical variables with respect to $t$,
we discretize a space with a mesh with a equal size $\Delta r$,
\beq
r_i=(i-1)\Delta r,~~~ i=1,...,N,
\eeq
and label a dynamical variable $F(t ,r _i)$ as $F_i^{(t )}$.
Here $F(t ,r _i)$ represents $\alpha,~a,~\bp,~\bxi$ and $\bpi$ collectively.
We choose $N=3001$.
A derivative with respect to $r ^2$ is approximated as
\bea
{\pa F\over\pa(r^2)}(r_i)&=&{1\over 2r_i}{\pa F\over\pa r}(r_i) \nn
&\approx&{1\over 2r_i}{F_{i+1}-F_{i-1}\over2\Delta r}
={F_{i+1}-F_{i-1}\over r^2_{i+1}-r^2_{i-1}}.
\eea
For each $F_i^{(t)}$ we integrate with respect to $t$ by the second 
order Runge-Kutta method, or the so-called ``predictor-corrector" method.

As for the boundary conditions of the center and the outer edge, we follow Hayley and Choptuik \cite{HC} as follows.
For $\bp,~\bxi$ and $\bpi$ at $r =0$, we employ a ``quadratic fit,"
\beq
F_1^{(t +\Delta t)}={4F_2^{(t +\Delta t)}-F_3^{(t +\Delta t)}\over 3}.
\eeq
For $\bxi$ and $\bpi$ at the outer boundary, we employ 
\bea
F_N^{(t +\Delta t)}&=&
\left({4F_N^{(t)}-F_N^{(t-\Delta t)}\over \Delta t}
+{4F_{N-1}^{(t+\Delta t)}-F_{N-2}^{(t-\Delta t)}\over \Delta r}\right)
\nn
&&\Big/\left({3\over\Delta t}+{3\over\Delta r}+{2\over r_N}\right)\ .
\eea

\begin{figure}
\psfig{file=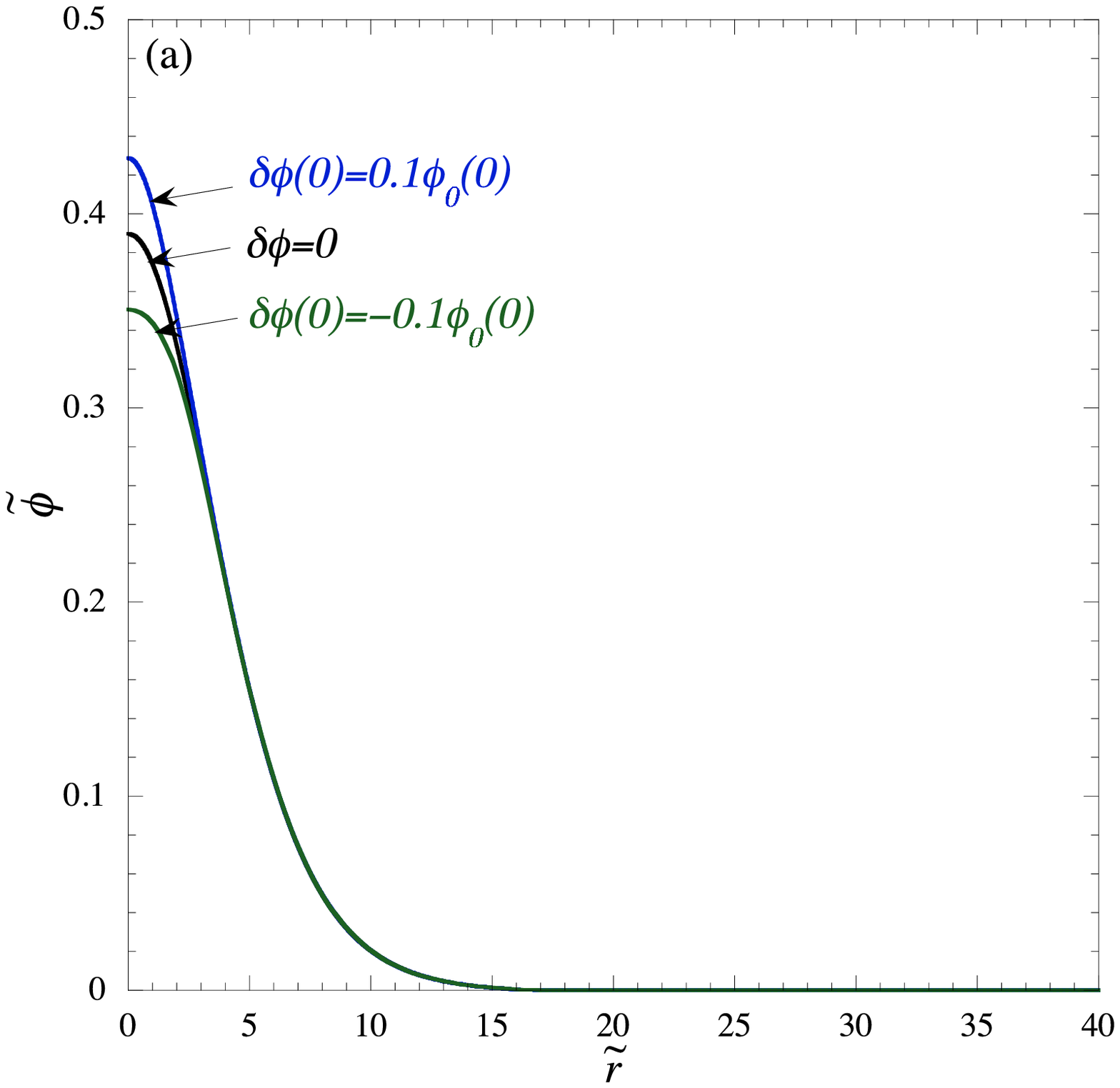,width=3in}
\psfig{file=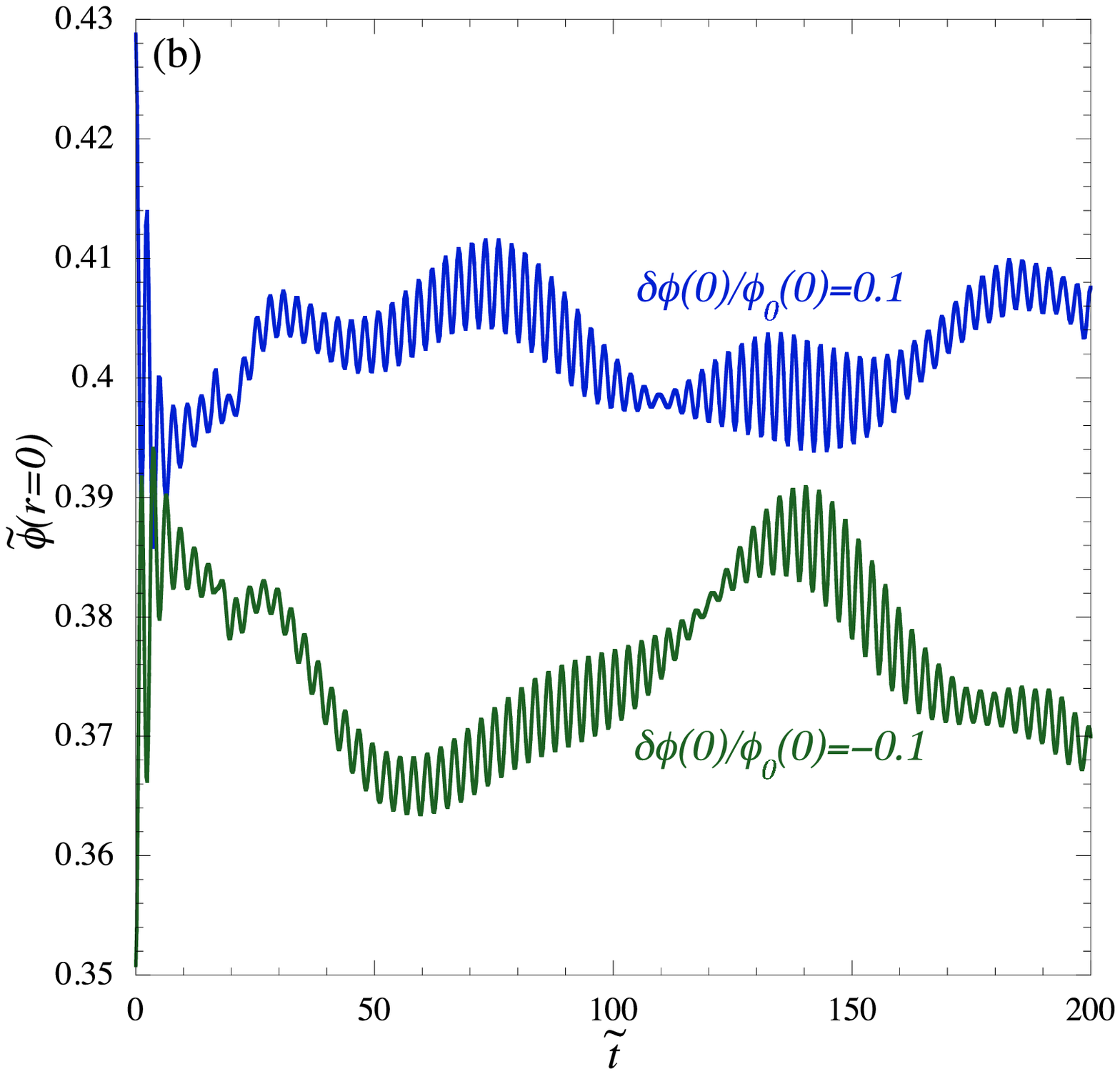,width=3in}
\caption{\label{A}
Perturbation of a stable solution $A$.
(a) Field configuration of the equilibrium solution and a perturbed initial configuration, 
which is given by (\ref{initial}) with $\delta\phi(0)/\phi_0(0)=0.1$ or -0.1 and $\tilde{L}=2$.
(b) Time-variation of $\tp(\tt,\tr=0)$.}
\end{figure}

To suppress numerical errors further, we apply numerical dissipation 
to $\bp_n,\bxi_n$ and $\bpi_n$, following Hayley and Choptuik \cite{HC}. 
After the next value $F_i^{(t +\Delta t)}$ is evaluated, we set
\bea
F_i^{(t +\Delta t)}\leftarrow F_i^{(t +\Delta t)}
-{\epsilon\over16}&(&F^{( t)}_{i+2}-4F^{( t)}_{i+1}+6F^{( t)}_{i}\nn
&&-4F^{( t)}_{i-1}+F^{( t)}_{i-2}).
\eea
where $\epsilon$ is an adjustable parameter in the range $0<\epsilon<1$, and we 
choose $\epsilon=0.5$.

\section{What happens if $Q$ is so large?}

\begin{figure}
\psfig{file=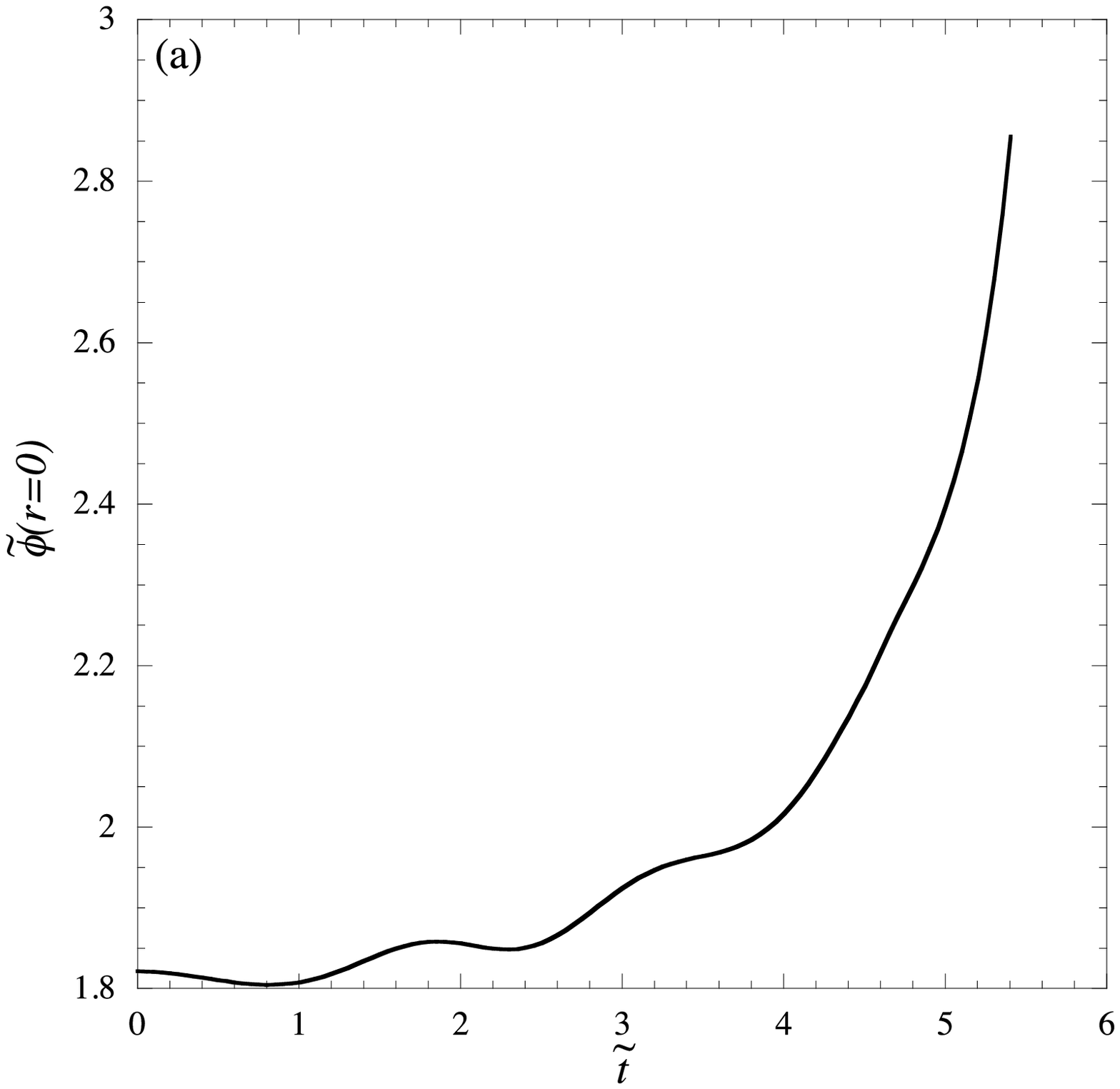,width=3in}
\psfig{file=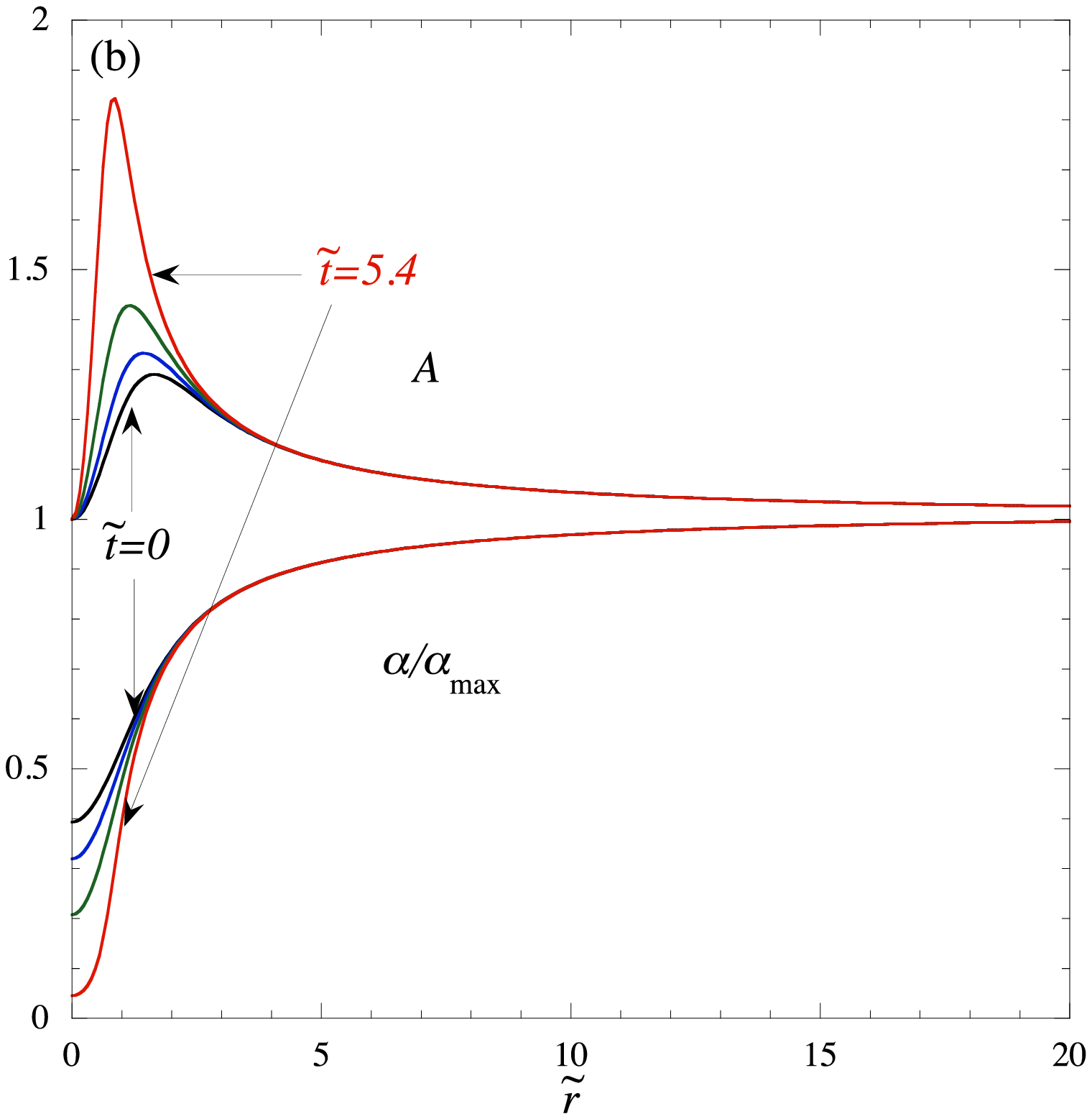,width=3in}
\caption{\label{B1}
Perturbation of an unstable solution $B$: $\delta\phi(0)/\phi_0(0)=0.01$ and $\tilde{L}=2$.
(a) Time-variation of $\tp(\tt,\tr=0)$.
(b) Snapshots of the metric functions $A$ and $\alpha$ at $\tt=0,~4,~5,~5.4$.}
\end{figure}

\begin{figure}
\psfig{file=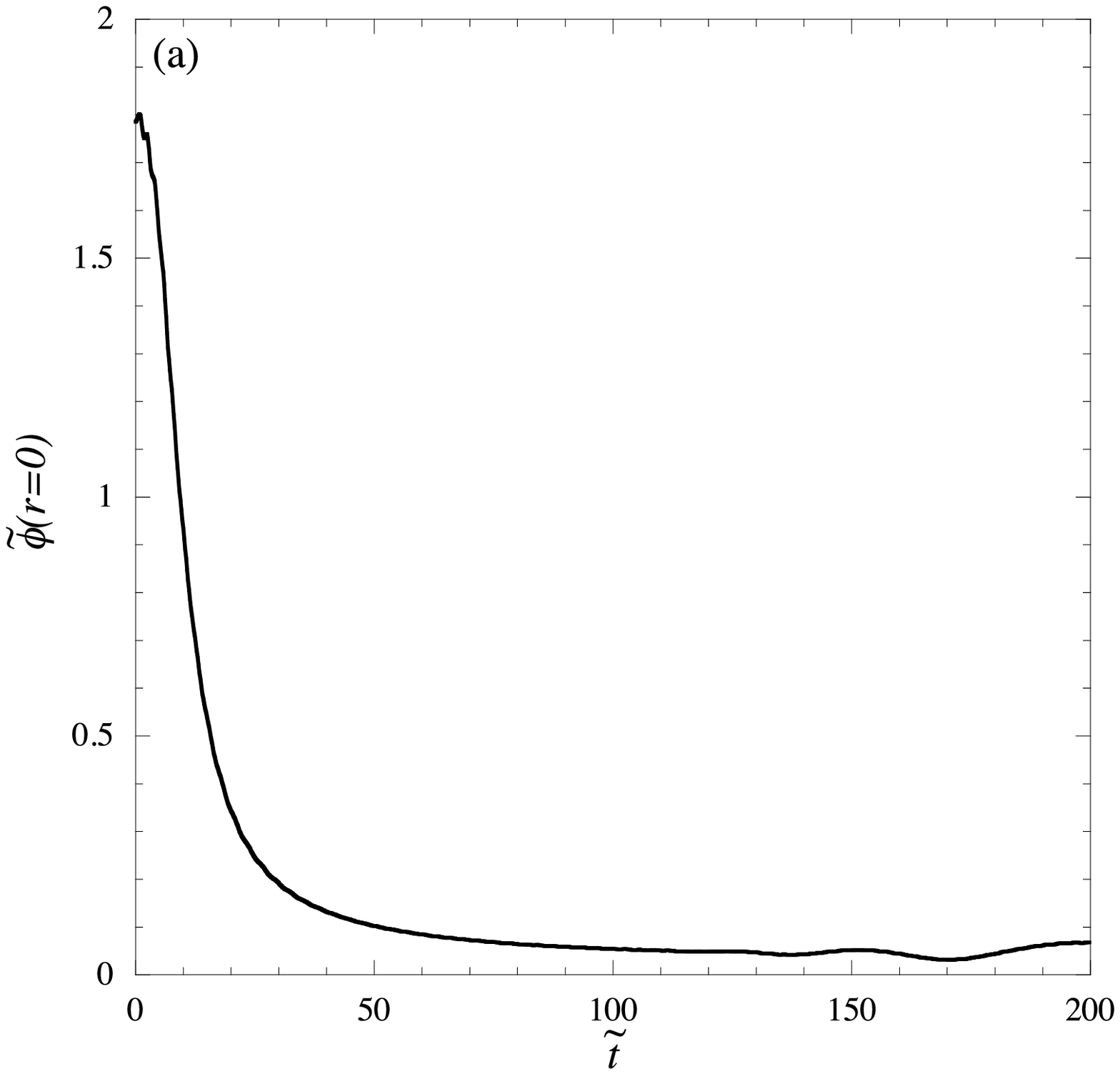,width=3in}
\psfig{file=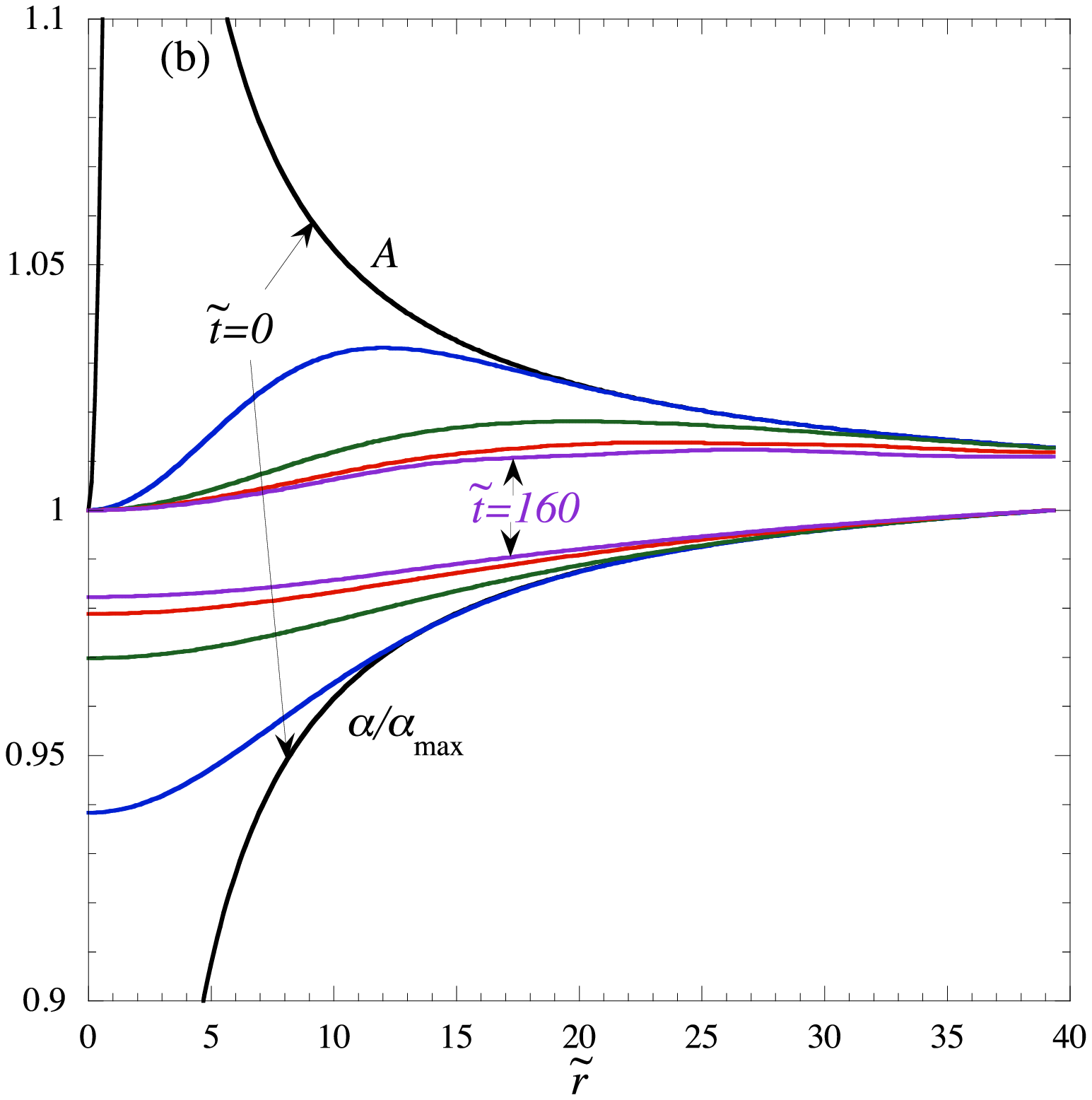,width=3in}
\caption{\label{B2}
Perturbation of an unstable solution $B$: $\delta\phi(0)/\phi_0(0)=-0.01$ and $\tilde{L}=2$.
(a) Time-variation of $\tp(\tt,\tr=0)$.
(b) Snapshots of the metric functions $A$ and $\alpha$ at $\tt=0,~40,~80,~120,~160$.}
\end{figure}

In the analysis method devised in the last section, we shall discuss what happens to Q-balls with the potential (\ref{AD}) if $Q$ is so large.
Because ordinary Q-balls exist only if  $K<0$, we concentrate on the case of $K<0$.
Specifically, we analyze dynamical field equations by giving perturbations to equilibrium solutions $A,~B,$ and $C$ in Fig.\ 3.

First, we consider perturbation of a stable solution $A$.
Figure 5(a) shows the field configuration of the equilibrium solution $A$ and 
perturbed initial configurations, which is given by (\ref{initial}) with $\delta\phi(0)/\phi_0(0)=0.1$ or -0.1 and $\tilde{L}=2$.
(b) shows the time-variation of $\phi(\tt,\tr=0)$, which indicates that the field continues to vibrate around the equilibrium configuration.
These results assure us that stable solutions indicated by energetics or by catastrophe theory are really stable.
The mean values of $\tp(t,0)$ for the two solutions are slightly different from each other because $Q$ is slightly changed by the perturbed field $\delta\phi$.

Secondly, we consider perturbation of an unstable solution $B$.
We give two types of perturbations.
Figure 6 shows the case of positive perturbation, $\delta\phi(0)/\phi_0(0)=0.01$.
(a) indicates that the field $\phi$ diverges in the center.
(b) tells us that $\alpha$ approaches to zero and $A$ diverges.
In the coordinate system (\ref{metric}) this behavior means a black hole is formed.
Figure 7 shows the case of negative perturbation, $\delta\phi(0)/\phi_0(0)=-0.01$.
We find that the Q-ball diffuses most of mass and charge but not all. It becomes a thick-wall Q-ball with much smaller charge.

\begin{figure}
\psfig{file=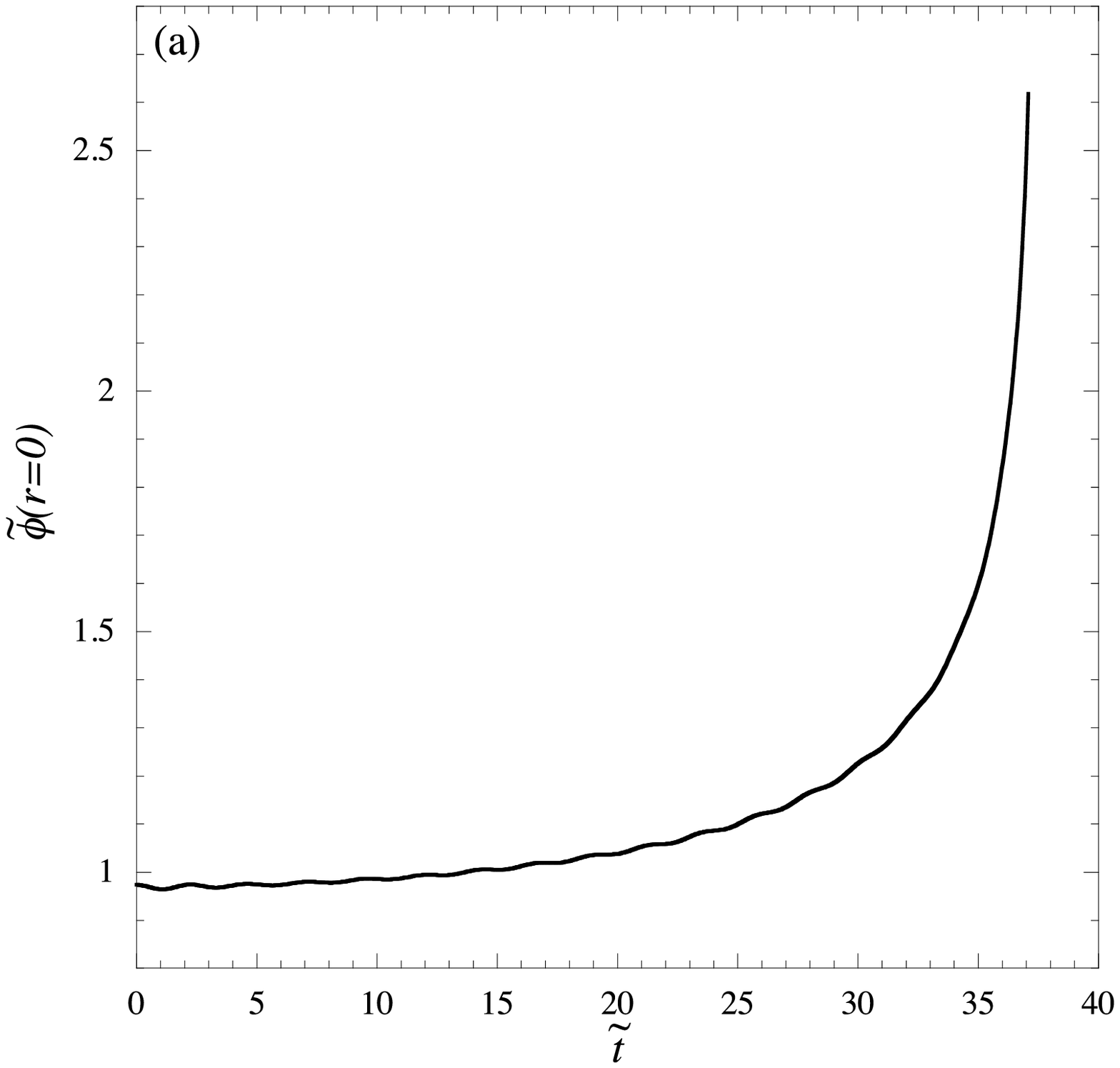,width=3in}
\psfig{file=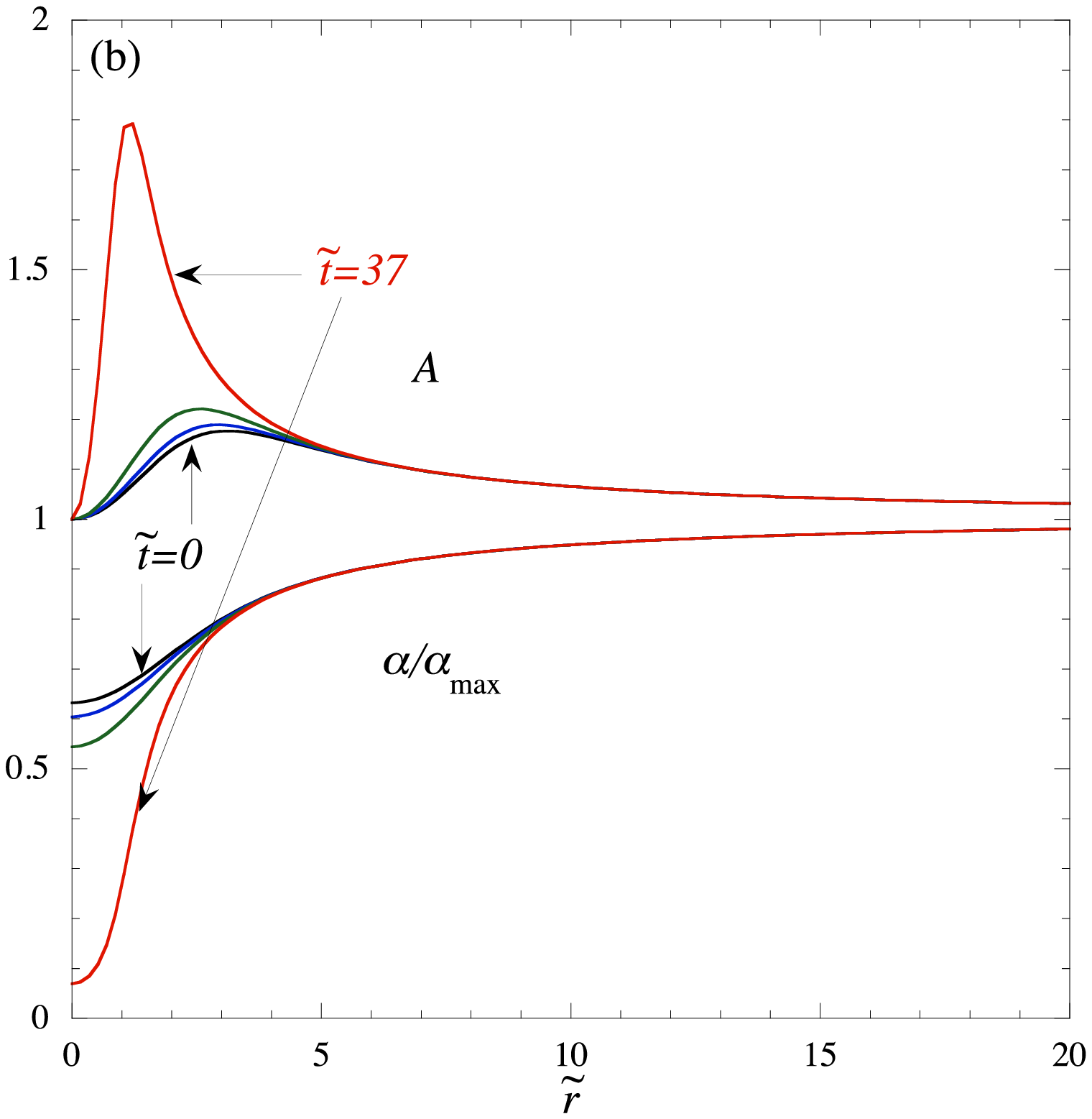,width=3in}
\caption{\label{C1}
Perturbation of the extremal solution $C$: $\delta\phi(0)/\phi_0(0)=0.01$ and $\tilde{L}=2$.
(a) Time-variation of $\tp(\tt,\tr=0)$.
(b) Snapshots of the metric functions $A$ and $\alpha$ at $\tt=0,~20,~30,~37$.}
\end{figure}

\begin{figure}
\psfig{file=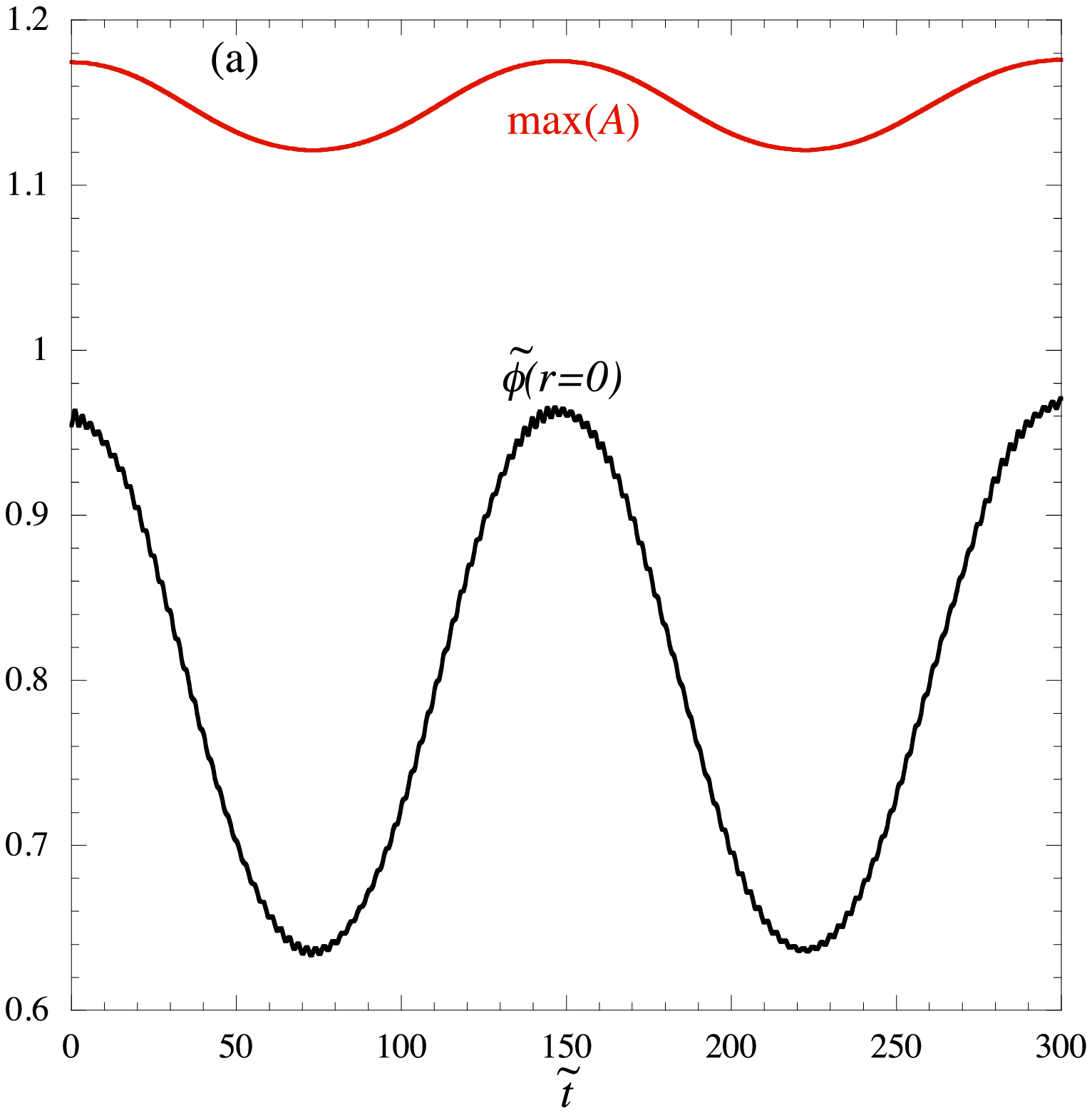,width=3in}
\psfig{file=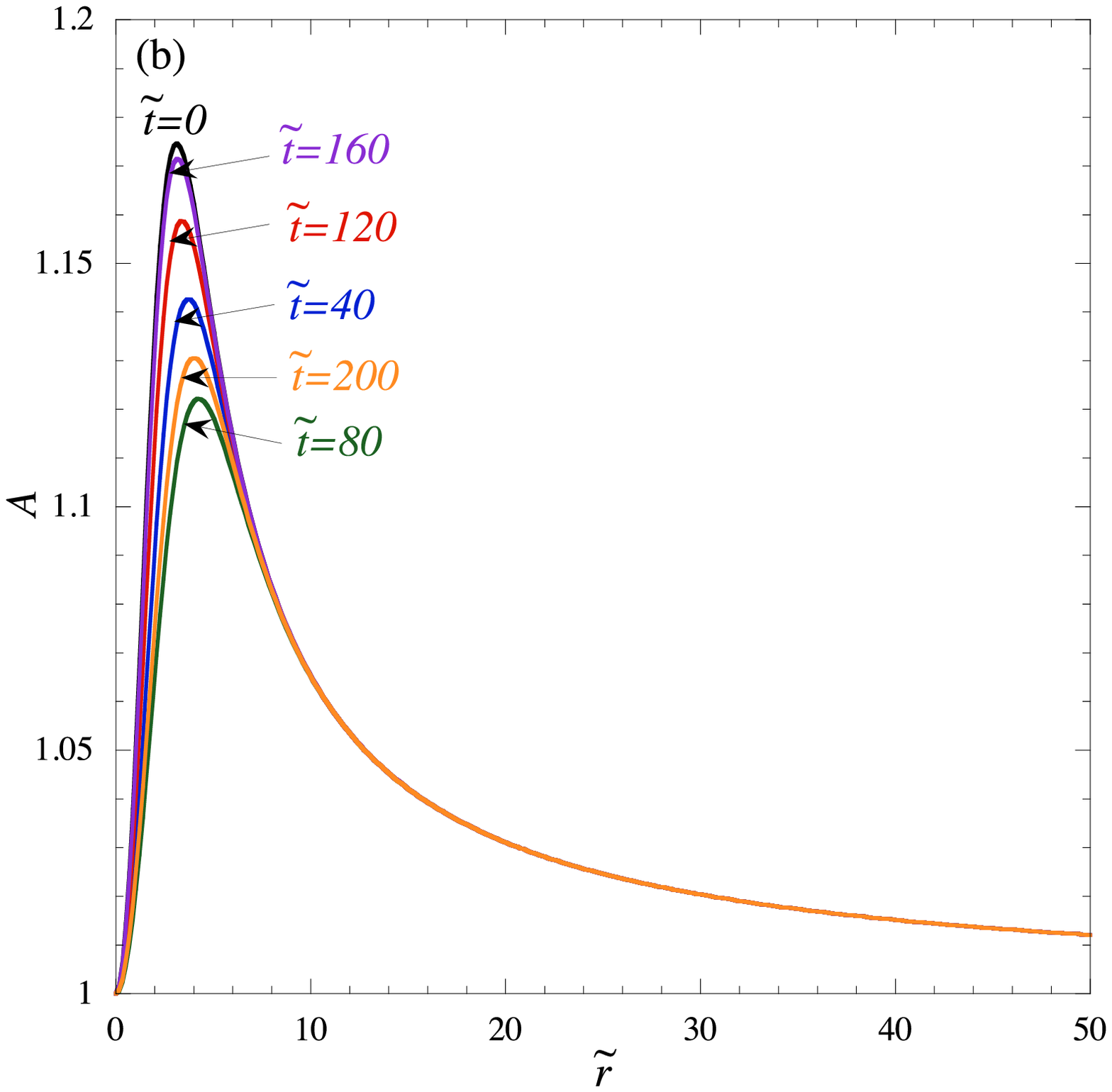,width=3in}
\caption{\label{C2}
Perturbation of the extremal solution $C$: $\delta\phi(0)/\phi_0(0)=-0.01$ and $\tilde{L}=2$.
(a) Time-variation of $\tp(\tt,\tr=0)$ and max$[A(\tt,\tr)]$.
(b) Snapshots of the metric functions $A$ at $\tt=0,~40,~80,~120,~160,~200$.}
\end{figure}

Thirdly, we consider perturbation of the extremal solution $C$ with $Q_{{\rm max}}$.
Again, we give two types of perturbations.
Figure 8 shows the case of positive perturbation, $\delta\phi(0)/\phi_0(0)=0.01$.
Like the unstable solution $B$, this Q-ball collapses and becomes a black hole.
Figure 9 shows the case of negative perturbation, $\delta\phi(0)/\phi_0(0)=-0.01$.
The behavior  in this situation is not analogous to that for $B$ in Fig.\ 7.
The Q-ball continues to oscillate without diffusing mass or charge.
The above results are also seen even if we take other values of $\tilde{L}$ and 
$\delta \phi (0)/\phi_0(0)$ as long as they are not so large.

Finally, we ascertain that the dynamics is virtually unchanged even if we choose $\kappa=10000$.
Figure 10 shows an example of the dynamical solutions, where except for $\kappa$ the parameters are the same as in Fig.\ 8.
We see that the Q-ball collapses and becomes a black hole in the same way as in Fig.\ 8.

Thus our numerical analysis has provided confirmation of our analytic argument that inflation cannot take place in the core of a Q-ball.
Furthermore, it indicates that the extremal solution and unstable solutions near it are critical solutions of black-hole formation \cite{Chop}.
In fact, this critical phenomenon was already found for mini-boson stars ($K=0$ in the model (\ref{AD}) by Hawley and Choptuik \cite{HC}.
It is reasonable that Q-balls with $K<0$ and with $K=0$ share the same property in the case that $Q=Q_{{\rm max}}$, or gravitational effects are so large.

\begin{figure}
\psfig{file=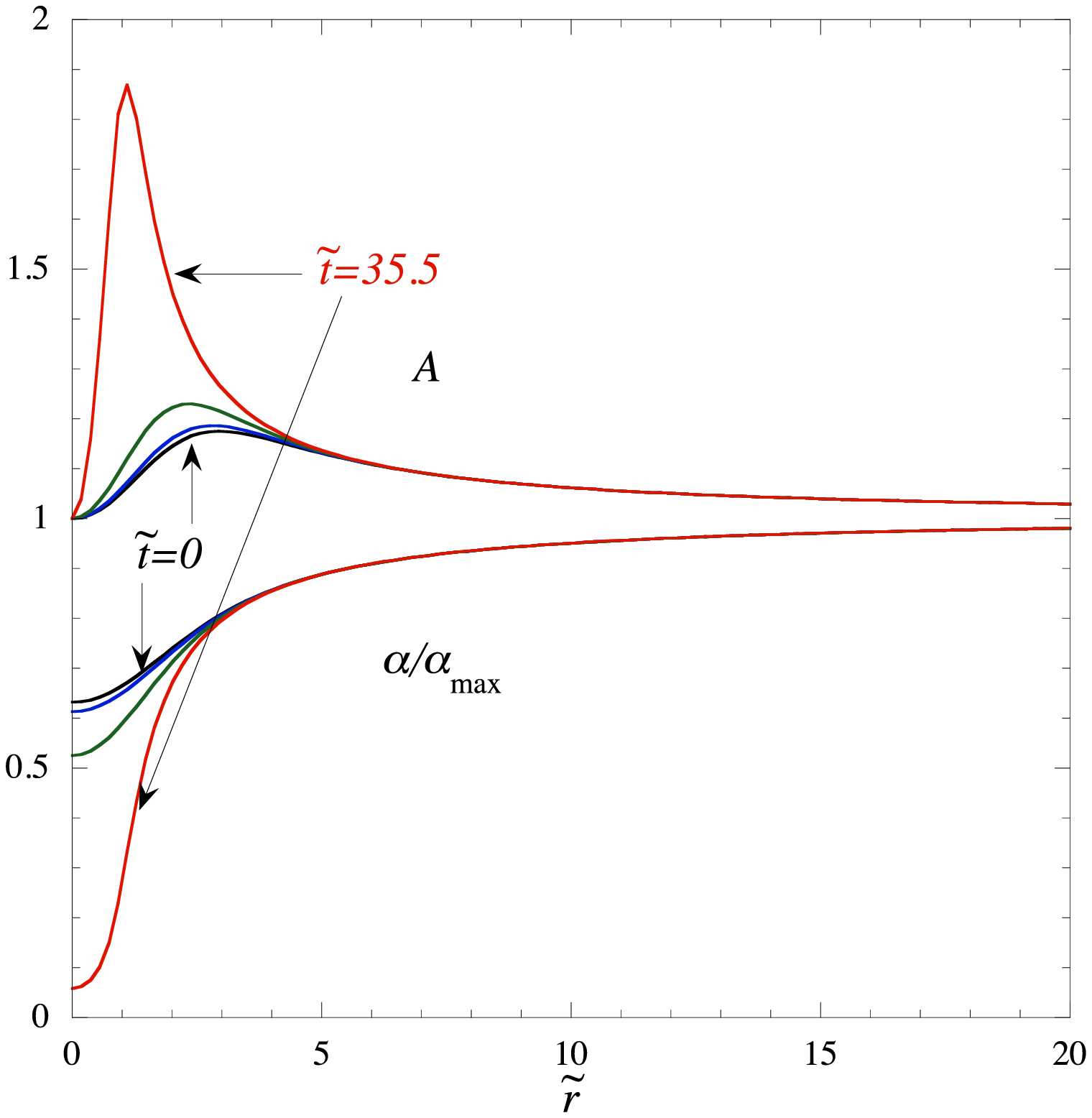,width=3in}
\caption{\label{C2}
Perturbation of the extremal solution for $\kappa=10000$ and $K=-0.01$.
We choose $\delta\phi(0)/\phi_0(0)=0.01$ and $\tilde{L}=2$.
The figure shows snapshots of the metric functions $A$ and $\alpha$ at $\tt=0,~20,~30,~35.5$.}
\end{figure}

\section{Concluding remarks}

We have addressed a question of what happens to Q-balls if $Q$ is close to $Q_{{\rm max}}$.
First, without specifying a model, we have shown analytically that the core of an equilibrium Q-ball has attractive nature and inflation cannot take place there.
Next, for the Affleck-Dine model, we have analyzed  perturbation of equilibrium solutions with $Q\approx Q_{{\rm max}}$ by numerical analysis of dynamical field equations.
We have found that the extremal solution with $Q=Q_{{\rm max}}$ and unstable solutions around it are ``critical solutions", which means the threshold of black-hole formation.

Specifically, for initial data (\ref{initial}) with (\ref{perturbation}), a black-hole is formed if $\delta\phi>0$.
If $\delta\phi<0$, a black hole is not formed, and there are two types of evolutions.
If the initial configuration is very close to the extremal Q-ball with $Q=Q_{{\rm max}}$, the Q-ball continues to oscillate without diffusing mass or charge.
In other cases,  the Q-ball diffuses most of mass and charge, and becomes a thick-wall Q-ball with much smaller charge.

Further study is necessary to understand detailed behavior of these critical solutions.
It is also interesting to investigate how such behavior of critical solutions depends on models $V(\phi)$.

\acknowledgements
This work was supported by Grant-in-Aid for Scientific Research on Innovative Areas No.\ 22111502.
The numerical calculations were carried out on SX8 at YITP in Kyoto University.


\begin{thebibliography}{99}
\bibitem{foot}In fact, because the potential (\ref{AD}) becomes negative for large $\phi$, 
there appears $\tilde{Q}_{\rm max}\sim10^{30}$.
However, if we introduce an unrenormalizable term in such a way that $V$ is nonnegative everywhere, 
such an upper limit disappears.
\end{thebibliography}
\end{document}